\documentclass[aps,prd,groupedaddress,showpacs,nofootinbib
]{revtex4}
\usepackage{graphicx,amssymb,bm}

\newcommand{\be}{\begin{equation}}
\newcommand{\ee}{\end{equation}}
\newcommand{\bea}{\begin{eqnarray}}
\newcommand{\eea}{\end{eqnarray}}
\newcommand{\beaa}{\begin{eqnarray*}}
\newcommand{\eeaa}{\end{eqnarray*}}

\newcommand{\nn}{\nonumber \\}
\newcommand{\e}{{\rm e}}

\begin{document}

\tolerance=5000

\title{Dark energy from modified $F(R)$-scalar-Gauss-Bonnet gravity}
\author{Shin'ichi Nojiri}
\email{nojiri@phys.nagoya-u.ac.jp}
\affiliation{Department of Physics, Nagoya University, Nagoya 464-8602. Japan}
\author{Sergei D. Odintsov\footnote{also at Lab. Fundam. Study, Tomsk State
Pedagogical University, Tomsk}}
\email{odintsov@ieec.uab.es}
\affiliation{Instituci\`{o} Catalana de Recerca i Estudis Avan\c{c}ats (ICREA)
and Institut de Ciencies de l'Espai (IEEC-CSIC),
Campus UAB, Facultat de Ciencies, Torre C5-Par-2a pl, E-08193 Bellaterra
(Barcelona), Spain}
\author{Petr V. Tretyakov}
\affiliation{Sternberg Astronomical Institute, Moscow 119992, Russia}

\begin{abstract}
The modified $F(R)$-scalar-Gauss-Bonnet gravity is proposed as dark energy
model. The reconstruction program for such theory is developed.
It is explicitly demonstrated that the known classical universe expansion
history (deceleration epoch, transition to acceleration and effective
quintessence, phantom or cosmological constant era) may naturally occur
in such unified theory for some (reconstructed) classes of scalar
potentials. Gauss-Bonnet assisted dark energy is also
proposed. The possibility of cosmic acceleration is studied there.

\end{abstract}


\maketitle

\section{Introduction}

It is quite possible that dark energy is some manifestation of the unknown
gravitational physics. The search of  the realistic gravitational
alternative for
dark energy (for a recent review, see \cite{review}) continues.
According to this approach, usual General Relativity which was valid at
deceleration epoch should be modified by
some gravitational terms which became relevant at current, accelerating
universe when curvature is getting smaller. Among the most popular
modified gravities which may successfully describe the cosmic speed-up is
$F(R)$ gravity. Very simple versions of such theory like $1/R$ \cite{CDTT}
and $1/R+R^2$ \cite{NO} may lead to the effective quintessence/phantom
late-time universe. Moreover, positive/negative powers of curvature in the
effective gravitational action may have stringy/M-theory
origin \cite{string}. Recently, big number of works was devoted to the
study of late-time cosmology and solar system tests checks in modified
$F(R)$ gravity \cite{FR,FR1}. While it is not easy to satisfy all known
solar system tests at once, it is possible to construct the explicit
models \cite{cap,lea} which describe the realistic universe expansion
history (radiation/matter dominance, transition to acceleration and
accelerating era).

Another theory proposed as gravitational dark energy is
scalar-Gauss-Bonnet gravity \cite{sasaki} which is closely related with
low-energy string effective action. The late-time universe
evolution and
comparison with astrophysical data in such
model was discussed in refs.\cite{sasaki,sami,neupane,koivisto}. The
possibility to extend such consideration to third order (curvature cubic)
terms in low-energy string effective action exists \cite{sami1}.
Moreover, one can develop the reconstruction program for such theories
(see \cite{recrev}, for a review). It has been demostrated \cite{rec} that
some scalar-Gauss-Bonnet gravities may be compatible with the known
classical history of the universe expansion.

In the present paper, we propose the unified
$F(R)$-scalar-Gauss-Bonnet gravity as dark energy model. The reconstruction
program for such model is explicitly developed. It is shown that it is
easier to realize the known, classical universe history (deceleration,
transition to acceleration and cosmic acceleration with effective $w$
close to $-1$) in such a model than in $F(R)$ or scalar-Gauss-Bonnet
gravity. Several different late-time regimes (effective quintessence,
phantom or cosmological constant) are investigated.
Moreover, it is indicated that it is possible to pass Solar System tests
in
such a unified model.
 It is shown also that
Gauss-Bonnet term may play an important role for other class of
gravitational models where it couples with scalar kinetic term.
The possibility of cosmic acceleration in such model is also demonstrated.

\section{Cosmic acceleration in $F(R)$-scalar-Gauss-Bonnet gravity}

In this section, let us introduce $F(R)$-scalar-Gauss-Bonnet gravity
whose action is given by
\be
\label{FGB1}
S=\int d^4 x \sqrt{-g}\left[ \frac{R}{2\kappa^2} + F(R) - \frac{\eta}{2}\partial_\mu \phi
\partial^\mu \phi - V(\phi) - \xi(\phi) G + L_{\rm matter}\right]\ .
\ee
Here $F(R)$ is a function of the scalar curvature $R$, $L_{\rm matter}$ is
matter Lagrangian density, and $G$ is  GB invariant:
$G=R^2 - 4 R_{\mu\nu} R^{\mu\nu} + R_{\mu\nu\rho\sigma}R^{\mu\nu\rho\sigma}$.

In (\ref{FGB1}), with $F(R)=0$ $\eta=1$ corresponds to
string-inspired scalar-Gauss-Bonnet gravity which has been
proposed as dark energy model \cite{sasaki} (it may be applied also for
resolution of the initial singularity in early universe \cite{ant}) and
$\eta=0$ to $f(G)$ gravity \cite{noplb}. Note that using trick of
 ref.\cite{NO} one can delete $F(R)$ term, at the price of the
introduction of second scalar. However, in such formulation the coupling
 (including derivatives) of new scalar with GB sector appears.

In the following, the metric is assumed to be in the spatially-flat FRW
form:
\be
\label{FGB3}
ds^2 = - dt^2 + a(t)^2 \sum_{i=1}^3 \left(dx^i\right)^2\ .
\ee
The FRW field equations  look as
\bea
\label{FGB4}
0&=& - \frac{R}{2\kappa^2} - F(R) + 6\left(H^2  + \dot H\right) \left(\frac{1}{2\kappa^2} + F'(R)\right)
 - 36 \left(4H^2 \dot H + H \ddot H\right)F''(R) \nn
&& + \frac{1}{2}{\dot\phi}^2 + V(\phi) + 24 H^3 \frac{d \xi(\phi(t))}{dt} + \rho_{\rm matter} \ ,\\
\label{FGB5}
0&=& \frac{R}{2\kappa^2} + F(R) - 2\left(\dot H + 3H^2\right)\left(\frac{1}{2\kappa^2} + F'(R) \right) \nn
&& + 48 \left(4H^2 \dot H + {\dot H}^2 + 2 H \ddot H\right)F''(R)
+ 72\left(4H\dot H + \ddot H\right)F'''(R) + \frac{1}{2}{\dot\phi}^2 - V(\phi) \nn
&& - 8H^2 \frac{d^2 \xi(\phi(t))}{dt^2} - 16H \dot H
\frac{d\xi(\phi(t))}{dt} - 16 H^3 \frac{d \xi(\phi(t))}{dt} + p_{\rm matter}\ .
\eea
and the scalar field equation is
\be
\label{FGB6}
0=\eta\left(\ddot \phi + 3H\dot \phi\right) + V'(\phi) + \xi'(\phi) G\ .
\ee
Here $R=12H^2 + 6\dot H$ and $G=24\left(\dot H H^2 + H^4\right)$

We now consider the perfect fluids with constant equation of state (EoS)
parameters
$w_i\equiv p_i / \rho_i$ as the matter. The energy density is
$\rho_i=\rho_{i0} a^{-3(1+w_i)}$ with a constant $\rho_{i0}$.
Let us parametrize the model with two proper functions $f(\phi)$ and
$g(t)$ as follows (compare with \cite{rec} for $F=0$ case),
\bea
\label{FGB9}
V(\phi) &=& \frac{\hat R}{2\kappa^2} + F(\hat R)
 - 6\left(g'\left(f(\phi)\right)^2 + g''\left(f(\phi)\right)\right) \left(\frac{1}{2\kappa^2} + F'(\hat R)\right) \nn
&& + 36\left(4g'\left(f(\phi)\right)^2g''\left(f(\phi)\right)
+ g'\left(f(\phi)\right)g'''\left(f(\phi)\right)\right)F''(\hat R)
 - 3g'\left(f(\phi)\right) \e^{g\left(f(\phi)\right)} U(\phi) \ ,\nn
\xi(\phi) &=& \frac{1}{8}\int^\phi d\phi_1 \frac{f'(\phi_1) \e^{g\left(f(\phi_1)\right)}
U(\phi_1)}{g'\left(f(\phi_1)\right)^2}\ ,\nn
U (\phi)&\equiv & \int^\phi d\phi_1 f'(\phi_1)\e^{-g\left(f(\phi_1)\right)}
\left(4 g''\left(f(\phi_1)\right) \left(\frac{1}{2\kappa^2} + F'(\hat R) \right) \right. \nn
&& + 12 \left(4g'\left(f(\phi_1)\right)^2g''\left(f(\phi_1)\right)
+ 4 g''\left(f(\phi_1)\right)^2 + 5 g'\left(f(\phi_1)\right)g'''\left(f(\phi_1)\right)\right) F''(\hat R) \nn
&& + 72 \left( 4 g'\left(f(\phi_1)\right) g''\left(f(\phi_1)\right)
+ g'''\left(f(\phi_1)\right) \right) F'''(\hat R) \nn
&& \left. + \frac{\eta}{f'(\phi_1 )^2} + \sum_ i (1+w_i)\rho_{i0}
a_0^{-3(1+w_i)}\e^{-3(1+w_i)g\left(f(\phi_1)\right)}\right) \ , \nn
\hat R &\equiv& 12 g'\left(f(\phi)\right)^2 + 6g''\left(f(\phi)\right) \ .
\eea
Here, the equations have the following solution:
\be
\label{FGB10}
a=a_0\e^{g(t)}\ \left(H= g'(t)\right)\ ,\quad \phi=f^{-1}(t)\ .
\ee
This may be considered as reconstruction of above modified gravity
from known universe history expansion.

In case $\eta=0$ in (\ref{FGB1}), we can freely redefine the scalar field
as $\phi\to \phi' = \phi'(\phi)$.
If $\phi$ depends on the time $t$ as $\phi=\phi(t)$, one can redefine the
scalar field properly and identify
the scalar field as $t$, $\phi=t$, that is $f(\phi)=\phi$. Then
Eq.(\ref{FGB9}) can be simplified as (for the case $F=0$ compare with \cite{recrev})
\bea
\label{FGB9B}
V(\phi) &=& \frac{\hat R}{2\kappa^2} + F(\hat R)
 - 6\left(g'\left(\phi\right)^2 + g''\left(\phi\right)\right) \left(\frac{1}{2\kappa^2} + F'(\hat R)\right) \nn
&& + 36\left(4g'\left(\phi\right)^2g''\left(\phi\right)
+ g'\left(\phi\right)g'''\left(\phi\right)\right)F''(\hat R)
 - 3g'\left(\phi\right) \e^{g\left(\phi\right)} U(\phi) \ ,\nn
\xi(\phi) &=& \frac{1}{8}\int^\phi d\phi_1 \frac{\e^{g\left(\phi_1\right)}
U(\phi_1)}{g'\left(\phi_1\right)^2}\ ,\nn
U (\phi)&\equiv & \int^\phi d\phi_1 \e^{-g\left(\phi_)\right)}
\left(4 g''\left(\phi_1\right) \left(\frac{1}{2\kappa^2} + F'(\hat R) \right) \right. \nn
&& + 12 \left(4g'\left(\phi_1\right)^2g''\left(\phi_1\right)
+ 4 g''\left(\phi_1\right)^2 + 5 g'\left(\phi_1\right)g'''\left(\phi_1\right)\right) F''(\hat R) \nn
&& + 72 \left( 4 g'\left(\phi_1\right) g''\left(\phi_1\right)
+ g'''\left(\phi_1\right) \right) F'''(\hat R) \nn
&& \left. + \sum_ i (1+w_i)\rho_{i0} a_0^{-3(1+w_i)}\e^{-3(1+w_i)g\left(\phi_1\right)}\right) \ , \nn
\hat R &\equiv& 12 g'\left(\phi\right)^2 + 6g''\left(\phi\right) \ .
\eea

As some example of $F(R)$, we consider \cite{CDTT}
\be
\label{FGB11}
F(R)=-\frac{\mu^4}{R}\ .
\ee
The original model has two solutions corresponding to deSitter universe, where $R$ is constant,
and asymptotic universe with effective $w=-2/3$. In the model
(\ref{FGB9}), one can realize any time developement
of the universe. For example, by choosing (\ref{FGB11}) and
\be
\label{FGB12}
g(t)=H_0 t + H_1\ln\left(\frac{t}{t_0}\right)\ ,
\ee
and $f(\phi)$ to be properly defined, we obtain
\be
\label{FGB13}
H(t)= H_0 + \frac{H_1}{t}\ .
\ee
When $t$ is small, $H$ (\ref{FGB13}) behaves as that in universe with perfect fluid with $w=-1 + 2/3H_1$ and
when $t$ is large, $H$ behaves as in deSitter space, where $H$ is a constant.
Then if we choose $H_1=2/3$, we find that before the acceleration epoch,
the universe behaves as matter dominated one
with $w=0$. After that, it enters to acceleration phase.
In the original model \cite{CDTT}, it was difficult to realize the matter dominated phase.
It is easy to see that matter dominance  phase (with subsequent
acceleration) can be easily realized by adding the Gauss-Bonnet term
(see Appendix for explicit form of scalar potentials).

As another example, we consider a model with (\ref{FGB11}) but
\be
\label{FGB14}
g(t)=\tilde H_0\ln\frac{t}{t_0} - \tilde H_1\ln\left(\frac{t_0 - t}{t_0}\right)\ ,
\ee
which gives
\be
\label{FGB15}
H(t)=\frac{\tilde H_0}{t} + \frac{\tilde H_1}{t_0 - t}\ .
\ee
Here $\tilde H_0$, $\tilde H_1$, and $t_0$ are positive constants.
When $t$ is small, $H$ (\ref{FGB15}) behaves in a way corresponding to the
perfect fluid with
$w=-1 + 2/3\tilde H_0$.
Then if we choose $\tilde H_0=2/3$, we find the matter dominated universe.
On the other hand, when $t\sim t_0$ is large, $H$ behaves as that in the phantom universe with $w=-1 - 2/3\tilde H_1<-1$ and
there will appear a big rip singularity at $t=t_0$.
The three-year WMAP data are analyzed in Ref.~\cite{Spergel}, which
show that the combined analysis of WMAP with the supernova Legacy
survey (SNLS) constrains the dark energy equation of state $w_{DE}$
pushing it clearly towards the cosmological constant value.
The marginalized best fit values of the equation of state parameter at 68$\%$ confidence
level are given by $-1.14\leq w_{DE} \leq -0.93$, which corresponds
to $\tilde H_1>10.7$ as $\tilde H_1$ is positive. In case one takes
as a prior that the universe is flat, the combined data gives
$-1.06 \leq w_{DE} \leq -0.90 $, which corresponds to $\tilde H_1>25.0$.
Therefore the possibility that $w_{DE}<-1$ has not been excluded.
As clear from (\ref{FGB9}) or (\ref{FGB9B}), the expressions of $V(\phi)$
 and $\xi_1(\phi)$ depend on the explicit form
of $F(R)$, say $\mu$ in (\ref{FGB11}), and the time-development of the universe $g(t)$, now $\tilde H_0$ and
$\tilde H_1$, (and $f(\phi)$ in case of (\ref{FGB9})).
Then the form of $F(R)$ is irrelevant to the WMAP data but only $g(t)$ is relevant as long as
we use the expressions in (\ref{FGB9}) or (\ref{FGB9B}).

One may also consider a model
\be
\label{FGB16}
g(t)=\hat H_0\ln\frac{t}{t_0} + \left(\hat H_1 - \hat H_0\right) \ln\left(\frac{t_0 + t}{t_0}\right)\ ,
\ee
with constants $\hat H_0$, $\hat H_1$, and $t_0$.
Then we obtain
\be
\label{FGB17}
H(t)=\frac{\hat H_0}{t} + \frac{\hat H_1 - \hat H_0}{t_0 + t}\ .
\ee
When $t$ is small, $H$ again behaves as $H\sim \hat H_0/t$ corresponding to the universe filled with perfect
fluid $w=-1 + 2/3\hat H_0$.
On the other hand, when $t\sim t_0$ is large, $H$ behaves as $H\sim \hat
H_1/t$ corresponding to the fluid
with $w=-1 + 2/3\hat H_1$.
Therefore with the choice $\hat H_0=2/3$, we find the matter dominated
universe in the early universe and with the choice
 $\hat H_1 > 1$, we obtain acceleratedly expanding universe, where
$w<-1/3$.
The constraint from only the WMAP data indicates $\hat H_1>21.4$ and that
from the combined data indicates $\hat H_1>15.0$.

One more example is $\Lambda$CDM-type cosmology:
\be
\label{LCDM1}
g(t)=\frac{2}{3(1+w)}\ln \left(\alpha \sinh \left(\frac{3(1+w)}{2l}\left(t - t_0 \right)\right)\right)\ ,\quad
\alpha^2\equiv \frac{1}{3}\kappa^2 l^2 \rho_0 a_0^{-3(1+w)}\ .
\ee
Here $l$ is the length scale given by cosmological constant $l\sim \left(10^{-33}\,{\rm eV}\right)^{-1}$
and $t_0$ is a constant. The time-development of the universe given by
 $g(t)$ (\ref{LCDM1}) can be
realized in the usual Einstein gravity with a cosmological constant $\Lambda$ and cold dark matter (CDM),
which could be regarded as  dust. Then in the present formulation, by
using $V(\phi)$ and $\xi_1(\phi)$ in
(\ref{FGB9}) or (\ref{FGB9B}), $\Lambda$CDM-type cosmology can be realized {\it without} introducing CDM as a matter.

One may also consider more general choice of $F(R)$ as in \cite{NO}
\be
\label{FGB18}
F(R)=-\frac{\alpha}{R}+\beta R^2 \ ,
\ee
or
\be
\label{FGB19}
F(R)=-\frac{\tilde\alpha}{R^n} + \tilde\beta R^m\ .
\ee
Even in this case, we can realize any time development of the universe.
For example, if we choose $g(t)$ as in (\ref{FGB12}),
$H$  (\ref{FGB13}) behaves as that in universe with perfect fluid with $w=-1 + 2/3H_1$ when $t$ is small
(especially by choosing $H_1=2/3$, the matter dominated phase occurs) and
$H$ behaves as in deSitter space when $t$ is large.
On the other hand, if we choose $g(t)$ as in (\ref{FGB14}), we obtain a model showing the transition from
the matter dominated phase to the phantom phase. Moreover, with the choice
of (\ref{FGB17}), the transition from the
matter dominated phase to the quintessence phase could be realized.

Let us now consider the string-inspired model\cite{sasaki}, where
\be
\label{FGB20}
V=V_0\e^{-\frac{2\phi}{\phi_0}}\ ,\quad \xi(\phi)=\xi_0 \e^{\frac{2\phi}{\phi_0}}\ ,
\ee
with constant parameters $V_0$, $\xi_0$, and $\phi_0$ (with $F(R)$  given by (\ref{FGB18})).
Different from the model (\ref{FGB9}), it is not straightforward to solve the FRW equations (\ref{FGB4}),
(\ref{FGB5}) and the scalar field equation
in this model and to find the behavior of the universe. In case
$\left| \frac{\alpha}{R}\right|,\ \left|\beta R^2\right| \ll
\left|\frac{R}{\kappa^2}\right|$,
the cosmology \cite{sasaki} could be reproduced, that is, if we choose,
\bea
\label{FGB22}
V_0 t_1^2&=& - \frac{1}{\kappa^2\left(1 + h_0\right)}\left\{3h_0^2 \left( 1 - h_0\right)
+ \frac{ \phi_0^2 \kappa^2 \left( 1 - 5 h_0\right)}{2}\right\}\ ,\nn
\frac{48 \xi_0 h_0^2}{t_1^2}&=& - \frac{6}{\kappa^2\left( 1 + h_0\right)}\left(h_0
   - \frac{ \phi_0^2 \kappa^2}{2}\right)\ .
\eea
we obtain a solution
\be
\label{FGB23}
H=\frac{h_0}{t}\ ,\quad \phi=\phi_0 \ln \frac{t}{t_1}\ ,
\ee
when $h_0>0$ or
\be
\label{FGB24}
H=-\frac{h_0}{t_s - t}\ ,\quad \phi=\phi_0 \ln \frac{t_s - t}{t_1}\ ,
\ee
when $h_0<0$ and a constant $t_1$.
We should also note that there could also be deSitter solution:
$H=H_0\ ,\quad \phi=p_0$.
Then the FRW equation (\ref{FGB4}) and the scalar field equation (\ref{FGB6}) have
the following simple forms:
\be
\label{FGB26}
0=-\frac{3}{\kappa^2}H_0^2 + \frac{\alpha}{8H_0^2} + V_0\e^{-2p_0/\phi_0}\ ,\quad
0=V_0\e^{-2p_0/\phi_0} - \xi_0\e^{2p_0/\phi_0}\ .
\ee
When $V_0$ and $\xi_0$ are positive, by combining two equations in (\ref{FGB26}), we obtain
\be
\label{FGB27}
0=-\left(\frac{3}{\kappa^2} - 2\sqrt{6V_0\xi_0}\right)H_0^2 + \frac{\alpha}{8H_0^2} \ .
\ee
If $\alpha$ is positive and $\frac{3}{\kappa^2} > 2\sqrt{6V_0\xi_0}$,
there can be a solution
$H_0^4=\frac{\alpha}{8\left(\frac{3}{\kappa^2} -
2\sqrt{6V_0\xi_0}\right)}$,
which gives
$\e^{2p_0/\phi_0}=\sqrt{\left(\frac{3}{\kappa^2} -
2\sqrt{6V_0\xi_0}\right)\frac{V_0}{3\alpha \xi_0}}$.
Therefore, in general, there could occur  deSitter solution as late-time
universe.

Hence, we demonstrated that in unified $F(R)$-scalar-GB gravity which may be
considered as string-inspired theory one may have the number of dark
energy scenarios of different type (effective quintessence, effective
cosmological constant or effective phantom). The corresponding classes of
scalar potentials may be easily constructed (see an explicit example of
such construction in the Appendix). It is interesting that in such unified
model it is easier to realize the known classical universe history (
radiation/matter dominance, decceleration-acceleration transition,
cosmic acceleration) than in pure $F(R)$ gravity \cite{cap,lea} or in
pure scalar-GB
gravity \cite{rec,recrev}. There is no problem to take into account usual matter
in such consideration. In this case it is even easier to reconstruct the
decceleration (radiation/matter dominance) phase before the acceleration epoch.
However, the explicit form of corresponding potentials is quite complicated.

\section{Gauss-Bonnet gravity assisted dark energy}

In this section we study modified gravity model motivated by gravity
assisted dark energy \cite{assist} where scalar kinetic term couples with
the function of Gauss-Bonnet invariant. It is different from
the model of previous section, but GB term plays again the important role here.
The starting action is:
\begin{equation}
S=\int d^4 x \sqrt{-g} \left[ \frac{1}{\kappa^2}R - f(G) L_d\right]\ ,
\label{1}
\end{equation}
where $L_d=-\frac{1}{2}g^{\mu\nu}\partial_{\mu}\phi\partial_{\nu}\phi$.
Taking the same FRW metric
and assuming $\phi$ only depends on time coordinate $t$, $\phi = \phi (t)$,
we find the solution of scalar field equation:
\begin{equation}
\dot \phi=f(G)^{-1}a^{-3} q\ .
\label{4}
\end{equation}
Here $q$ is a constant.

The FRW field equation is found to be:
\be
\label{6g}
\frac{6}{\kappa^2}H^2-\frac{q^2}{a^6f(G)} \left[\frac{1}{2}+12H^2(\dot H+7H^2)\frac{f'(G)}{f(G)}
+24^2\left(\frac{f'(G)}{f(G)}\right)^2H^4(H\ddot H+4\dot H H^2+2\dot H^2)-12H^3 \frac{\frac{d}{dt}f'(G)}{f(G)} \right]=0 \ .
\ee
Here $f'\equiv df(G)/dG$.
The following choice of $f(G)$ is convenient: $f(G)=\left(\frac{G}{\mu^4} \right)^{\alpha}$.
Here $\alpha$ (non-integer) and $\mu^4$ are constants.
Then in order that $f(G)$ is real, we only consider the case that $G$ has a definite sign and when $G$ is positive (negative),
$\mu^4>0$ ($\mu^4<0$).
One can see that for $\alpha=-1$, FRW equation takes the very simple form:
\be
\label{6k}
\frac{H^2}{\kappa^2}+\frac{288 q^2H^4}{\mu^4a^6}=0\ .
\ee
When $\mu^4>0$, the equation has only trivial solution $H=0$.
When $\mu^4<0$,  the following solution exists:
\be
\label{6kb}
a(t)=2^{2/3}\left(-\frac{2q^2}{\mu^4}\right)^{1/6}t^{1/3}\ .
\ee
which gives $H=1/3t$ and $G=-16/27t^4$.

Now for general $\alpha$, by using a constant $x$, we assume
\be
\label{6l}
a=a_0t^x\ ,\quad \left(H=\frac{x}{t}\right)\ .
\ee
Then it follows $x=\frac{2\alpha+1}{3}$,
$a_0^6=48q^2\kappa^2\left\{\frac{81\mu^4}{(2\alpha+ 1)(\alpha-1)}\right\}^\alpha
\frac{(\alpha^2-3\alpha-1)}{(\alpha-1)}$.
Since $G=\frac{16(2\alpha+1)(\alpha -1)}{27t^4}$,
we find $\mu^4>0$ when $\alpha>1$ or $\alpha<-1/2$ and $\mu^4<0$ when $-1/2<\alpha<1$.
Since $\ddot a =\frac{2}{9}a_0(2\alpha+1)(\alpha-1)t^{\frac{2\alpha-5}{3}}$
the universe accelerates ($\ddot a>0$) if $\alpha>1$ or $\alpha<-1/2$.
We should also note that $81\mu^4/(2\alpha+ 1)(\alpha-1)$ is positive by assumption.
Since $a_0^6>0$, one gets $(3-\sqrt{13})/2<\alpha<1$ or $\alpha>(3+\sqrt{13})/2$.
Hence, if $(3-\sqrt{13})/2<\alpha<1$, the expansion of the universe
is decelerating but
if $\alpha>(3+\sqrt{13})/2$, the expansion is accelerating.

The effective equation of state parameter in our case is $w=\frac{1-2\alpha}{1+2\alpha}$.
If $\alpha<-\frac{1}{2}$ the effective phantom era occurs, while
if $\alpha>1$ the effective quintessence era emerges.

Let us add the perfect
fluid $p=(\gamma-1)\rho$, where $\gamma\in(0,2)$, to our system.
The FRW equation may be presented in the following form:
\begin{equation}
\frac{H^2}{\kappa^2}=\rho+\rho_{G}\ , \label{8}
\end{equation}
where
\be
\label{8b}
\rho_G=\frac{48q^2\mu^{4\alpha}}{a^6(\dot HH^2+H^4)^{\alpha+2}}\left[\alpha(\alpha+1)
\ddot H H^5+2(2\alpha^2+6\alpha+1)\dot HH^6+2(\alpha+1)\left(\alpha+\frac{1}{2}\right)\dot H^2H^4+(1+7\alpha)H^8
\right]\ .
\ee
The  energy density $\rho$ satisfies the equation $\dot\rho=-3\gamma H\rho$.
 From (\ref{8}), it follows $\rho=H^2/\kappa^2-\rho_G$. Substituting it
into the conservation law, one gets: $2\frac{\dot
HH}{\kappa^2}-\dot \rho_G +3\gamma \frac{H^3}{\kappa^2} -3\gamma
H\rho_G=0$. By assuming (\ref{6l}), we find $a=a_0
t^{\frac{1+2\alpha}{3}}$ as in the case without the perfect fluid
and $\rho=\rho_0 t^{-\gamma(1+2\alpha)}$. If $\rho\propto\rho_G$,
we find $\gamma(1+2\alpha)=2$. For the case that matter is dust
with $\gamma=1$, if $\alpha>\frac{1}{2}$, the matter could
dominate in the early universe ($t\to 0$) and $\rho_G$ dominates
in the late universe $t\to +\infty$. Hence, if
$\alpha>(3+\sqrt{13})/2$, there could occur the transition from
matter dominated phase to the accelerated expansion. It is
interesting to note that de Sitter solution is impossible in this
scenario with only  $\rho_G$.

Thus, we demonstrated the possibility  to realize the
Gauss-Bonnet gravity assisted dark energy with late-time cosmic acceleration.
The decceleration era may naturally emerge in such scenario before the acceleration with
subsequent transition to acceleration.

\section{Discussion}

In summary, we studied late-time, dark energy era in $F(R)$-scalar-Gauss-Bonnet gravity.
The reconstruction method for such model is developed.
It is shown that it may be reconstructed from the known universe expansion
history so that for some class of scalar potentials the radiation/matter
dominance is realized subsequently transiting to cosmic acceleration.
Moreover, any type of cosmic speed-up: the effective quintessence, phantom
or $\Lambda$-CDM era may occur after deceleration epoch.
It is remarkable that it is easier to achieve the deceleration era
transiting to dark era than in $F(R)$ or scalar-Gauss-Bonnet gravity.

Gauss-Bonnet gravity assisted dark energy is proposed. It is shown
that cosmic acceleration may naturally occur in such theory as well.
It is interesting that adding the scalar potential to kinetic term in such
theory one can construct the Gauss-Bonnet induced model for the dynamical
origin of the
effective cosmological constant in close analogy with ref.\cite{assist}.

Let us discuss now solar system tests for our model.
It has been argued recently that $F(R)$-gravity does not pass the solar
system tests \cite{Chiba}.
Recently, however, in \cite{HS}, the conditions that even pure
$F(R)$-gravity (of special form) could satisfy the solar system
and cosmological tests are derived. The conditions are:
\be
\label{HS2}
\lim_{R\to\infty} F (R) = \mbox{const}\ ,\quad
\lim_{R\to 0} F(R) = 0\ .
\ee
An explicit example of such function (with specific values of parameters)
is  given by
\be
\label{HS1}
F(R)=-\frac{m^2 c_1 \left(R/m^2\right)^n}{c_2 \left(R/m^2\right)^n + 1}\ .
\ee
Hence, we can  choose $F(R)$ as (\ref{HS1}) and then apply the
formulations (\ref{FGB9}) (for $\eta=1$ case)
and (\ref{FGB9B}) (for $\eta=0$ case) and therefore
we can realize arbitrary (accelerating) cosmology by properly choosing
$V(\phi)$ and $\xi(\phi)$. In this case, our theory passes solar system
tests.
For $F(G)$-gravity, which corresponds to $\eta=0$ case,
 it has been shown that the scalar field $\phi$ does
not propagate and the only propagating mode is graviton,
at least in the de Sitter background.
The model proposed in this paper is the hybrid of $F(R)$-gravity
and the scalar-Gauss-Bonnet or $F(G)$-gravity.
Hence, the above arguments  indicate that cosmological and
solar
system tests
could be satisfied at once if we choose $F(R)$ (as above), $V(\phi)$, and
$\xi(\phi)$, properly.  Of course, to satisfy all cosmological tests some
tuning of parameters of our theory may be necessary what should be checked
explicitly for any particular model.

\section*{Acknowledgements}

We thank M. Sasaki for useful discussions.
The investigation by S.N. has been supported in part by the
Ministry of Education, Science, Sports and Culture of Japan under
grant no.18549001
and 21st Century COE Program of Nagoya University
provided by Japan Society for the Promotion of Science (15COEG01),
and that by S.D.O. has been supported in part  by the
projects FIS2006-02842, FIS2005-01181
(MEC,Spain), by the project 2005SGR00790 (AGAUR,Catalunya), by LRSS
projects N4489.2006.02 and N1157.2006.02 and by RFBR grants 05-02-17450, 06-01-00609 (Russia).

\appendix

\section{}

In this Appendix for the simple cosmology case  (\ref{FGB11}) and (\ref{FGB12}),
we present the explicit form of $V(\phi)$ and $\xi_1$.

For the $F(R)$-scalar-Gauss-Bonnet model ($\eta=1$ in (\ref{FGB11})),
we further choose $f(\phi)=f_0\phi$. Then
$V(\phi)$ and $\xi_1$  (\ref{FGB9}) have the following form:
\bea
\label{AFGB9}
V(\phi) &=& \frac{\hat R}{2\kappa^2} - \frac{\mu^4}{12\left(H_0 + \frac{H_1}{f_0\phi}\right)^2 - \frac{6H_1}{f_0^2\phi^2}}
 - 6\left(\left(H_0 + \frac{H_1}{f_0\phi}\right)^2 - \frac{H_1}{f_0^2\phi^2}\right)
\left(\frac{1}{2\kappa^2} + \frac{\mu^4}{\left(12\left(H_0 + \frac{H_1}{f_0\phi}\right)^2
 - \frac{6H_1}{f_0^2\phi^2}\right)^2}\right) \nn
&& + 72\left( - 4\left(H_0 + \frac{H_1}{f_0\phi}\right)^2 \frac{H_1}{f_0^2 \phi^2}
+ 2\left(H_0 + \frac{H_1}{f_0\phi}\right)\frac{H_1}{f_0^3 \phi^3}\right)\frac{\mu^4}{
\left(12\left(H_0 + \frac{H_1}{f_0\phi}\right)^2 - \frac{6H_1}{f_0^2\phi^2}\right)^3} \nn
&& -3 \left(H_0 + \frac{H_1}{f_0\phi}\right)\e^{H_0f_0\phi}\left(\frac{f_0\phi}{t_0}\right)^{H_1} U(\phi) \ ,\nn
\xi_1(\phi) &=& \frac{1}{8}\int^\phi d\phi_1 \frac{f_0 \e^{H_0f_0\phi}\left(\frac{f_0\phi}{t_0}\right)^{H_1}
U(\phi_1)}{\left(H_0 + \frac{H_1}{f_0\phi}\right)^2}\ ,\nn
U (\phi)&\equiv &
\int^\phi d\phi_1 f_0\e^{-H_0f_0\phi_1}\left(\frac{f_0\phi_1}{t_0}\right)^{-H_1}
\left(-4 \frac{H_1}{f_0^2\phi_1^2}
 \left(\frac{1}{2\kappa^2}
+ \frac{\mu^4}{\left(12\left(H_0 + \frac{H_1}{f_0\phi_1}\right)^2 - \frac{6H_1}{f_0^2\phi_1^2}\right)^2} \right) \right. \nn
&& - 24 \left(- 4\left(H_0 + \frac{H_1}{f_0\phi}\right)^2 \frac{H_1}{f_0^2\phi^2}
+ 4 \frac{H_1^2}{f_0^4\phi_1^4}
+ 10 \left(H_0 + \frac{H_1}{f_0\phi_1}\right)\frac{H_1}{f_0^3 \phi^3}
\right)
\frac{\mu^4}{\left(12\left(H_0 + \frac{H_1}{f_0\phi}\right)^2 - \frac{6H_1}{f_0^2\phi^2}\right)^3} \nn
&& + 432 \left( - 4 \left(H_0 + \frac{H_1}{f_0\phi}\right)\frac{H_1}{f_0^2\phi^2}
+ \frac{H_1}{f_0^3 \phi^3} \right) \frac{\mu^4}{\left(12\left(H_0 + \frac{H_1}{f_0\phi}\right)^2
 - \frac{6H_1}{f_0^2\phi^2}\right)^4} \nn
&& \left. + \frac{\eta}{f_0^2} + \sum_ i (1+w_i)\rho_{i0}
a_0^{-3(1+w_i)}
\e^{-3(1+w_i)H_0f_0\phi}\left(\frac{f_0\phi}{t_0}\right)^{-3(1+w_i)H_1}
\right) \ .
\eea
This defines the class of scalar potentials which lead to
above cosmological solution.
For $f(G)$ gravity  with $\eta=0$, the similar form of $V(\phi)$
and $\xi(\phi)$ may be derived.


\begin{thebibliography}{99}

\bibitem{review}
S.~Nojiri and S.~D.~Odintsov,
Int.\ J.\ Geom.\ Meth.\ Mod.\ Phys.\ {\bf 4}, 115 (2007)
[arXiv:hep-th/0601213].

\bibitem{CDTT}
S.~Capozziello,
Int.\ J.\ Mod.\ Phys.\ D {\bf 11}, 483 (2002);
S.~Capozziello, S.~Carloni and A.~Troisi,
arXiv:astro-ph/0303041;
S.~M.~Carroll, V.~Duvvuri, M.~Trodden and S.~Turner,
Phys.\ Rev.\ D {\bf 70} (2004) 043528.

\bibitem{NO}
S.~Nojiri and S.~D.~Odintsov,
Phys.\ Rev.\ D {\bf 68}, 123512 (2003)
[arXiv:hep-th/0307288].

\bibitem{string}
S.~Nojiri and S.~D.~Odintsov,
Phys.\ Lett.\ B {\bf 576}, 5 (2003)
[arXiv:hep-th/0307071].

\bibitem{FR}
F.~Faraoni,
arXiv:gr-qc/0607116;
arXiv:gr-qc/0511094;
M.~Ruggiero and L.~Iorio,
arXiv:gr-qc/0607093;
A.~Cruz-Dombriz and A.~Dobado,
arXiv:gr-qc/0607118;
N.~Poplawski,
arXiv:gr-qc/0607124;
A.~Brookfield, C.~van~de~Bruck and L.~Hall,
arXiv:hep-th/0608015;
Y.~Song, W.~Hu and I.~Sawicki,
arXiv:astro-ph/0610532;
B.~Li, K.~Chan and M.~Chu,
arXiv:astro-ph/0610794;
X.~Jin, D.~Liu and X.~Li,
arXiv:astro-ph/0610854;
T.~Sotiriou and S.~Liberati,
arXiv:gr-qc/0604006;
T.~Sotiriou,
arXiv:gr-qc/0611107;
arXiv:gr-qc/0604028;
R.~Bean, D.~Bernat, L.~Pogosian, A.~Silvestri and M.~Trodden,
arXiv:astro-ph/0611321;
I.~Navarro and K.~Van~Acoleyen,
arXiv:gr-qc/0611127;
A.~Bustelo and D.~Barraco,
arXiv:gr-qc/0611149;
G.~Olmo,
arXiv:gr-qc/0612047;
T.~Saidov and A.~Zhuk,
arXiv:hep-th/0612227;
arXiv:hep-th/0604131;
F.~Briscese, E.~Elizalde, S.~Nojiri and S.~D.~Odintsov,
Phys.\ Lett.\ B {\bf 646}, 105 (2007)
[arXiv:hep-th/0612220];
L.~Amendola, R.~Gannouji, D.~Polarski and S.~Tsujikawa,
arXiv:gr-qc/0612180;
S.~Baghram, M.~Farhang and S.~Rahvar,
arXiv:astro-ph/0701013;
D.~Bazeia, B.~Carneiro~da~Cunha, R.~Menezes and A.~Petrov,
arXiv:hep-th/0701106;
P.~Zhang,
arXiv:astro-ph/0701662;
B.~Li and J.~Barrow,
arXiv:gr-qc/0701111;
S.~Bludman,
arXiv:astro-ph/0702085;
I.~Sawicki and W.~Hu,
arXiv:astro-ph/0702278;
T.~Rador,
arXiv:hep-th/0702081;
arXiv:hep-th/0701267;
L.~Sokolowski,
arXiv:gr-qc/0702097;
V.~Faraoni,
arXiv:gr-qc/0703044;
arXiv:astro-ph/0610734;
S.~Rahvar and Y.~Sobouti,
arXiv:astro-ph/0704.0680;
O.~Bertolami, C.~Boehmer, T.~Harko and F.~Lobo,
arXiv:gr-qc/0704.1733 .

\bibitem{FR1}
S.~Nojiri and S.~Odintsov,
Gen.\ Rel.\ Grav.\ {\bf 36}, 1765 (2004)
[arXiv:hep-th/0308176];
P.~Wang and X.~Meng,
arXiv:astro-ph/0406455;
arXiv:gr-qc/0311019;
M.~Abdalla, S.~Nojiri and S.~D.~Odintsov,
Class.\ Quant.\ Grav.\ {\bf 22}, L35 (2005)
[arXiv:hep-th/0409177];
G.~Cognola, E.~Elizalde, S.~Nojiri, S.~D.~Odintsov and S.~Zerbini,
JCAP {\bf 0502}, 010 (2005)
[arXiv:hep-th/0501096];
S.~Capozziello, V.~Cardone and A.~Troisi,
arXiv:astro-ph/0501426;
G.~Allemandi, A.~Borowiec, M.~Francaviglia and S.~D.~Odintsov,
arXiv:gr-qc/0504057;
G.~Allemandi, M.~Francaviglia, M.~Ruggiero and A.~Tartaglia,
arXiv:gr-qc/0506123;
T.~Multamaki and I.~Vilja,
arXiv:astro-ph/0506692;
J.~A.~R.~Cembranos,
Phys.\ Rev.\ D {\bf 73}, 064029 (2006)
[arXiv:gr-qc/0507039];
T.~Koivisto and H.~Kurki-Suonio,
arXiv:astro-ph/0509422;
T.~Clifton and J.~Barrow,
arXiv:gr-qc/0509059;
O.~Mena, J.~Santiago and J.~Weller,
arXiv:astro-ph/0510453;
M.~Amarzguioui, O.~Elgaroy, D.~Mota and T.~Multamaki,
arXiv:astro-ph/0510519;
I.~Brevik,
arXiv:gr-qc/0601100;
R.~Woodard,
arXiv:astro-ph/0601672;
T.~Koivisto,
arXiv:astro-ph/0602031;
S.~Perez Bergliaffa,
arXiv:gr-qc/0608072;
T.~Faulkner, M.~Tegmark, E.~Bunn and Y.~Mao,
arXiv:astro-ph/0612569;
G.~Cognola, M.~Castaldi and S.~Zerbini,
arXiv:gr-qc/0701138;
S.~Capozziello and R.~Garattini,
arXiv:gr-qc/0702075.

\bibitem{cap}
S.~Nojiri, S.~D.~Odintsov,
Phys.\ Rev.\ D {\bf 74}, (2006) 086005
[arXiv:hep-th/0608008];
hep-th/0610164;
S.~Capozziello, S.~Nojiri, S.~D.~Odintsov and A.~Troisi,
Phys.\ Lett.\ B {\bf 639}, 135 (2006).

\bibitem{lea}
S.~Fay, S.~Nesseris and L.~Perivolaropoulos,
arXiv:gr-qc/0703006;
S.~Fay, R.~Tavakol and S.~Tsujikawa,
arXiv:astro-ph/0701479.

\bibitem{sasaki}
S.~Nojiri, S.~D.~Odintsov and M.~Sasaki,
Phys.\ Rev.\ D {\bf 71}, 123509 (2005)
[arXiv:hep-th/0504052].

\bibitem{sami}
M.~Sami, A.~Toporensky, P.~Tretjakov and S.~Tsujikawa,
Phys.\ Lett.\ B {\bf 619}, 193 (2005)
[arXiv:hep-th/0504154];
G.~Calcagni, S.~Tsujikawa and M.~Sami,
Class.\ Quant.\ Grav.\ {\bf 22}, 3977 (2005)
[arXiv:hep-th/0505193];
S.~Tsujikawa and M.~Sami,
arXiv:hep-th/0608178;
Z.~Guo, N.~Ohta and S.~Tsujikawa,
arXiv:hep-th/0610336;
A.~Sanyal,
arXiv:astro-ph/0608104.

\bibitem{neupane}
B.~Leith and I.~Neupane,
arXiv:hep-th/0702002;
B.~Carter and I.~Neupane,
JCAP {\bf 0606}, 004 (2006)
[arXiv:hep-th/0512262].

\bibitem{koivisto}
T.~Koivisto and D.~Mota,
arXiv:hep-th/0609155;
arXiv:astro-ph/0606078;
T.~Sotiriou and E.~Barausse,
arXiv:gr-qc/0612065.

\bibitem{sami1}
S.~Nojiri, S.~D.~Odintsov and M.~Sami,
Phys.\ Rev.\ D {\bf 74}, 046004 (2006)
[arXiv:hep-th/0605039];
D.~Konikowska and M.~Olechowski,
arXiv:hep-th/0704.1234;
E. Elizalde, S. Jhingan, S. Nojiri, S.D. Odintsov, M. Sami and I.
Thongkool,
arXiv:0705.1211[hep-th].

\bibitem{recrev}
S.~Nojiri and S.~D.~Odintsov,
arXiv:hep-th/0611071.

\bibitem{rec}
G.~Cognola, E.~Elizalde, S.~Nojiri, S.~D.~Odintsov and S.~Zerbini,
arXiv:hep-th/0611198.

\bibitem{ant}
I.~Antoniadis, J.~Rizos and K.~Tamvakis,
Nucl.\ Phys.\ B {\bf 415}, 497 (1994);
P.~Kanti, J.~Rizos and K.~Tamvakis,
Phys.\ Rev.\ D {\bf 59}, 083512 (1999);
N.~Mavromatos and J.~Rizos,
Phys.\ Rev.\ D {\bf 62}, 124004 (2000);
S.~Kawai, M.~Sakagami and J.~Soda,
Phys.\ Lett.\ B {\bf 437}, 284 (1998);
S.~Kawai and J.~Soda,
Phys.\ Rev.\ D {\bf 59}, 063506 (1999).

\bibitem{noplb}
S.~Nojiri and S.~D.~Odintsov,
Phys.\ Lett.\ B {\bf 631}, 1 (2005)
[arXiv:hep-th/0508049];
G.~Cognola, E.~Elizalde, S.~Nojiri, S.~D.~Odintsov and S.~Zerbini,
Phys.\ Rev.\ D {\bf 73}, 084007 (2006)
[arXiv:hep-th/0601008].

\bibitem{Spergel}
D.~N.~Spergel {\it et al.},
arXiv:astro-ph.0603449.

\bibitem{assist}
S.~Nojiri and S.~D.~Odintsov,
Phys.\ Lett.\ B {\bf 599}, 137 (2004)
[arXiv:astro-ph/0403622];
T.~Inagaki, S.~Nojiri and S.~D.~Odintsov,
JCAP {\bf 0506}, 010 (2005)
[arXiv:gr-qc/0504054].

\bibitem{Chiba}
A.~L.~Erickcek, T.~L.~Smith, M.~Kamionkowski,
Phys.\ Rev.\ D {\bf 74}, 121501 (2006)
[arXiv:astro-ph/0610483];
T.~Chiba, T.~L.~Smith, A.~L.~Erickcek,
arXiv:astro-ph/0611867.


\bibitem{HS}
W.~Hu and I.~Sawicki,
arXiv:0795.1158.

\end{thebibliography}
\end{document}